\documentstyle[amstex,preprint,aps]{revtex}

\begin{document}
\title{Thermodynamic properties of ferromagnetic/superconductor/ferromagnetic
nanostructures}
\author{I. Baladi\'{e} and A. Buzdin}
\address{Condensed Matter Theory Group, CPMOH, UMR 5798, Universit\'{e} Bordeaux 1,
33405 Talence Cedex,\\
France}
\date{10/09/2002}
\maketitle

\begin{abstract}
The theoretical description of the thermodynamic properties of
ferromagnetic/superconductor/ferromagnetic (F/S/F) systems of nanoscopic
scale is proposed. Their superconducting characteristics strongly depend on
the mutual orientation of the ferromagnetic layers. In addition, depending
on the transparency of S/F interfaces, the superconducting critical
temperature can exhibit four different types of dependences on the thickness
of the F-layer. The obtained results permit to give some practical
recommendations for the spin-valve effect experimental observation. In this
spin-valve sandwich, we also expect a spontaneous transition from parallel
to anti-parallel ferromagnetic moment orientation, due to the gain in the
superconducting condensation energy.
\end{abstract}

\section{Introduction}

The peculiar character of the proximity effect in
superconducting/ferromagnet (S/F) systems is due to the strong exchange
field acting on the electrons in the ferromagnet and provoking the
oscillatory-like behavior of the superconducting order parameter. Several
interesting phenomena inherent to the S/F hybrid structures have been
predicted and subsequently observed on experiments: non-monotonous
dependance of the critical temperature in S/F structures on the thickness of
the ferromagnetic layer $\cite{kuprianov90,Radovic91,Tagirov Physica
C,Proshin,Jiang85,Muihge,Lazar}$, $\pi $-junction realization in S/F/S
systems $\cite{Panyukov82,Kuprianov91,ryazanov,Kontos2}$ and local
quasiparticle density of states oscillation in S/F\ structures$\cite
{buzdin00,baladie,kontos}$.

In recent years, a great progress has been achieved in preparation of high
quality hybrid S/F\ systems, especially high quality interfaces, which could
be quite interesting for possible applications. In particular, a very
promising system is the F/S/F spin-valve sandwich, where
spin-orientation-dependant superconductivity has been predicted in $\cite
{Ryzhanova99,baladie2,Tagirov}.$

In this article, we present the results of detailed theoretical studies of
the properties of F/S/F systems containing a thin superconducting layer
(compared to the superconducting coherence length). We analyze the influence
of the F-layer thickness and the S/F interface transparency on the
spin-valve superconductivity effect. The last part of the article is devoted
to the thermodynamic properties of the spin-valve : we calculate the
superconducting order parameter and the superconducting condensation energy
for parallel and anti-parallel spin orientation of the F-layers. We discuss
also the possibility of a spontaneous phase transition, by decreasing the
temperature, from parallel to anti-parallel spin orientation.

\section{General equations}

We will concentrate on the studies of the properties of an F/S/F trilayer
system with F-layers of thickness $d_{f}$ and an S-layer of thickness $%
d_{s}, $ see Fig. 1. Assuming that the dirty limit conditions are held in
all layers, we may use the complete set of Usadel equations \cite{Usadel70}
in the superconducting layer and in the F-layers. In the superconducting
layer the Usadel Green functions $F$ and $G$ satisfy

\begin{equation}
-\frac{D_{s}}{2}\overrightarrow{\bigtriangledown }\left[ G(x,\omega )%
\overrightarrow{\bigtriangledown }F\left( x,\omega \right) -F(x,\omega )%
\overrightarrow{\bigtriangledown }G\left( x,\omega \right) \right] +\omega
F(x,\omega )=\Delta (x)G(x),  \label{Usadel2}
\end{equation}
in F-layers they verify \cite{baladie} 
\begin{equation}
\left( \omega +ih\right) F\left( x,\omega \right) -\frac{D_{f}}{2}%
\overrightarrow{\bigtriangledown }\left[ G(x,\omega )\overrightarrow{%
\bigtriangledown }F\left( x,\omega \right) -F(x,\omega )\overrightarrow{%
\bigtriangledown }G\left( x,\omega \right) \right] =0,  \label{Usadel ferro}
\end{equation}
and in both layers

\begin{equation}
G^{2}\left( x,\omega \right) +F(x,\omega )F^{\ast }(x,-\omega )=1,
\label{autocoherence}
\end{equation}
where $D_{s}$ and $D_{f}$ are the diffusion coefficients in S and F-layer
respectively, $\omega =2\pi T\left( n+1/2\right) $ are the Matsubara
frequencies and $h(x)$ is the exchange field in the F-layers. In the case of
the parallel orientation of the magnetization of the F-layers, the exchange
field is $h(x)=h$ for $x<-d_{s}/2$ and $x>d_{s}/2$ whereas in the
anti-parallel case $h(x)=h$ for $x>d_{s}/2$ and $h(x)=-h$ for $x<-d_{s}/2.$
The Usadel equations are completed by the self-consistency equation in the
form \cite{Abrikosov} 
\begin{equation}
\Delta \ln \frac{T}{T_{c}}+\pi T\sum_{\omega }\left( \frac{\Delta }{\left|
\omega \right| }-F_{s}\right) =0,  \label{Auto-cohérence générale}
\end{equation}
and by the boundary conditions at the S/F\ boundaries \cite{Kuprianov88} 
\begin{gather}
\frac{\partial F_{s}}{\partial x}=\gamma \frac{\partial F_{f}}{\partial x},
\label{Conditions aux limites} \\
F_{s}=F_{f}\pm \xi _{f}\gamma _{B}\frac{\partial F_{f}}{\partial x}, 
\nonumber
\end{gather}
where $\gamma =\frac{\sigma _{s}}{\sigma _{f}},$ $\sigma _{f}$ $\left(
\sigma _{s}\right) $ is the conductivity of the F-layer $\left( \text{%
S-layer above }T_{c}\right) ,$ $\xi _{f}=\sqrt{\frac{D_{f}}{2h}}$, $\xi _{s}=%
\sqrt{\frac{D_{s}}{2T_{c}}}$ is the superconducting coherence length of the
S-layer, the parameter $\gamma _{B}=\frac{R_{b}\sigma _{f}}{\xi _{f}},$
where $R_{b}$ is the S/F boundary resistance per unit area. In the second
boundary condition, the sign before the spatial derivative of $F_{f}$
depends on the relative orientation of the $x$-axis and the normal of the
ferromagnet surface. If the normal is parallel to the $x$-axis $\left(
x=d_{s}/2\right) $ the minus sign is required, in the other case $\left(
x=-d_{s}/2\right) $ the positive sign is required. The parameter $\gamma
_{B} $ is directly related to the transparency of the interface $T=\frac{1}{%
1+\gamma _{B}}$ \cite{AArts}. The limit $T=0$ $\left( \gamma _{B}=\infty
\right) $ corresponds to a vanishingly small boundary transparency, and the
limit $T=1$ $\left( \gamma _{B}=0\right) $ corresponds to a perfectly
transparent interface. At the interface between the vacuum and the
ferromagnet, the boundary condition is simply written as $\frac{\partial
F_{f}}{\partial x}=0.$

\section{Usadel equations for thin superconducting interlayer}

The mutual influence of superconductivity and ferromagnetism reveals
interesting effects for S-layer thickness smaller or of the order of
magnitude of the superconducting coherence length $\xi _{s}$, otherwise, we
have practically independent bulk superconductor and ferromagnetic systems.
In addition, the case $d_{s}\ll \xi _{s}$ has an analytical solution, that
is the reason why we will suppose this condition to be satisfied in the
following analysis$.$ In this limit, the small spatial variations of the
Green functions in S-layer can be taken into account by a simple expansion
to the order $x^{2}$

\begin{eqnarray}
F_{s} &=&F_{0}\left( 1+\alpha x+\frac{\beta }{2}x^{2}\right) ,
\label{expansion of green function} \\
G_{s} &=&G_{0}\left( 1+ax+\frac{b}{2}x^{2}\right) ,
\end{eqnarray}
where $F_{0}$ and $G_{0}$ are the values of the anomalous and normal Green
functions at the center of the S-layer. Using $\left( \ref{Usadel2}\right) $
and $\left( \ref{autocoherence}\right) ,$ we finally obtain an effective
Usadel equation for thin superconducting layers

\begin{equation}
\left[ \omega -\frac{D_{s}\beta }{2G_{0}}-\frac{D_{s}\alpha ^{2}F_{0}^{2}}{%
4G_{0}^{3}}\right] F_{0}=\Delta G_{0}.  \label{Usadel linearisé général}
\end{equation}
The coefficients $\alpha $ and $\beta $ in expression $\left( \ref{Usadel
linearisé général}\right) $ have to be found by using the boundary
conditions at the F/S interfaces. As it may be easily demonstrated from the
boundary conditions, the term containing $\alpha $ is by a factor $\left(
d_{s}/\xi _{s}\right) \ll 1$ smaller than the term with $\beta ,$
consequently this term can be neglected. Thus, in our approximation of thin
S-layer, the Usadel equations take the following simple form

\begin{eqnarray}
\left[ \omega -\frac{D_{s}\beta }{2G_{0}}\right] F_{0} &=&\Delta G_{0},
\label{Usadel et beta} \\
F_{0}^{2}+G_{0}^{2} &=&1,
\end{eqnarray}
where the coefficient $\beta $ plays the role of a pair-breaking parameter.
The boundary conditions on the function $F_{s}$, following from $\left( \ref
{expansion of green function}\right) ,$ are

\begin{eqnarray}
\left( F_{s}^{\prime }/F_{s}\right) _{-d_{s}/2} &=&\alpha -d_{s}\beta /2,
\label{frpimes sur fs} \\
\left( F_{s}^{\prime }/F_{s}\right) _{d_{s}/2} &=&\alpha +d_{s}\beta /2. 
\nonumber
\end{eqnarray}
By adding and substracting the previous equations we can find the
coefficients $\alpha $ and $\beta $ from the boundary conditions on $F_{s}.$
It is easy to demonstrate that the ratios $\left( F_{s}^{\prime
}/F_{s}\right) _{-d_{s}/2}$ and $\left( F_{s}^{\prime }/F_{s}\right)
_{d_{s}/2}$ are directly related to the corresponding ratios in the
ferromagnet, using the boundary conditions $\left( \ref{Conditions aux
limites}\right) $

\begin{equation}
\left( F_{s}^{\prime }/F_{s}\right) _{\pm d_{s}/2}=\frac{\gamma \left(
F_{f}^{\prime }/F_{f}\right) _{\pm d_{s}/2}}{1\mp \xi _{f}\gamma _{B}\left(
F_{f}^{\prime }/F_{f}\right) _{\pm d_{s}/2}}.
\label{fprime condition aux limites}
\end{equation}
In the next section, we will determine the critical temperature of the
S-layer under general transparency conditions at the S/F interfaces. In a
second part we will study the thermodynamics of the F/S/F structure at
arbitrary temperature, in the limit of high and low transparencies.

\section{Spin orientation dependance of the critical temperature}

Close to the critical temperature and assuming that the exchange field in
the ferromagnet is sufficiently strong $\left( h\gg T_{c}\right) $, the
Usadel equation in the ferromagnet can be simplified as

\begin{equation}
-\frac{\partial ^{2}F_{f}(x,\omega )}{\partial x^{2}}+\frac{2ihsgn\left(
\omega \right) }{D_{f}}F_{f}(x,\omega )=0
\end{equation}
Using the boundary condition at a vacuum interface, we readily find the
following solution for the Usadel Green functions in the parallel case (the
upperscript $P$ refers to the parallel case) for positive $\omega $ (the
case $\omega <0$ is obtained by making the substitution $k_{n}\rightarrow
k_{n}^{\ast }$)

\begin{eqnarray}
F_{f}^{P}\left( x>d_{s}/2\right) &=&A\cosh k_{n}\left[ x-\left(
d_{f}+d_{s}/2\right) \right] ,  \nonumber \\
F_{f}^{P}\left( x<-d_{s}/2\right) &=&B\cosh k_{n}\left[ x+\left(
d_{f}+d_{s}/2\right) \right] ,
\end{eqnarray}
with $k_{n}=\left( 1+i\right) \sqrt{\frac{h}{D_{f}}}$. Analogously for the
anti-parallel case (the upperscript $A$ refers to the anti-parallel case)

\begin{eqnarray}
F_{f}^{A}\left( x>d_{s}/2\right) &=&C\cosh k_{n}\left[ x-\left(
d_{f}+d_{s}/2\right) \right] ,  \nonumber \\
F_{f}^{A}\left( x<-d_{s}/2\right) &=&D\cosh k_{n}^{\ast }\left[ x+\left(
d_{f}+d_{s}/2\right) \right] ,
\end{eqnarray}
These solutions give immediately the value of the ratios $\left(
F_{f}^{\prime }/F_{f}\right) _{\pm d_{s}/2}$ and consequently, see $\left( 
\ref{fprime condition aux limites}\right) $ the ratios $\left( F_{s}^{\prime
}/F_{s}\right) _{\pm d_{s}/2}$

\begin{eqnarray}
\left( F_{s}^{\prime }/F_{s}\right) _{d_{s}/2} &=&-\frac{\gamma k_{n}\tanh
\left( k_{n}d_{f}\right) }{1+\xi _{f}k_{n}\gamma _{B}\tanh \left(
k_{n}d_{f}\right) },  \nonumber \\
\left( F_{s}^{\prime }/F_{s}\right) _{-d_{s}/2}^{P} &=&-\left( F_{s}^{\prime
}/F_{s}\right) _{d_{s}/2},  \label{dérivée de f dans supra} \\
\left( F_{s}^{\prime }/F_{s}\right) _{-d_{s}/2}^{A} &=&\frac{\gamma
k_{n}^{\ast }\tanh \left( k_{n}^{\ast }d_{f}\right) }{1+\xi _{f}k_{n}^{\ast
}\gamma _{B}\tanh \left( k_{n}^{\ast }d_{f}\right) }.  \nonumber
\end{eqnarray}
Then, with the help of $\left( \ref{frpimes sur fs}\right) ,$ we may easily
obtain the pair-breaking parameter $\beta .$ Close to $T_{c},$ the Usadel
equation may be linearized over $F_{0}$ and the normal Green function is $%
G_{0}=sgn\left( \omega \right) $, thus the equation $\left( \ref{Usadel
linearisé général}\right) $ is simply written in first order of $F_{0}$ as 
\begin{equation}
\left[ \left| \omega \right| -\frac{D_{s}\beta }{2}\right] F_{0}=\Delta .
\end{equation}
Using this relation and the self-consistency equation $\left( \ref
{Auto-cohérence générale}\right) $ we can write down the expression for the
critical temperature of the S-layer in the following general form 
\begin{equation}
\ln \frac{T_{c}}{T_{c0}}=\Psi \left( \frac{1}{2}\right) -%
\mathop{\rm Re}%
\Psi \left\{ \frac{1}{2}+\frac{1}{\tau \left( d_{f}\right) \pi T_{c}}%
\right\} ,  \label{temperature critique}
\end{equation}
where $T_{c0}$ is the critical temperature of the S-layer without any
proximity effect. This type of expression reminds the corresponding formula
for the critical temperature of a superconductor with magnetic impurities $
\cite{Abrikosov-Gorkov}$, though the ''magnetic scattering time'' $\tau $
may be complex in our system. It is easy to verify that in the parallel
case, the effective magnetic scattering rate $\tau ^{-1}$ is indeed complex
and given by the expression 
\begin{equation}
\tau ^{P}\left( d_{f}\right) ^{-1}=\tau _{0}^{-1}\frac{(1+i)\tanh \left( 
\widetilde{d_{f}}\right) }{1+\widetilde{\gamma _{B}}\tanh \left( \widetilde{%
d_{f}}\right) },
\end{equation}
and $\tau ^{-1}$\ in the anti-parallel case is real

\begin{equation}
\tau ^{A}\left( d_{f}\right) ^{-1}=%
\mathop{\rm Re}%
\left( \tau ^{P}\left( d_{f}\right) ^{-1}\right) ,
\end{equation}
here $\widetilde{d_{f}}=\left( 1+i\right) \frac{d_{f}}{\sqrt{2}\xi _{f}}$, $%
\widetilde{\gamma _{B}}=\left( 1+i\right) \frac{\gamma _{B}}{\sqrt{2}}$ and $%
\tau _{0}^{-1}=\frac{\gamma T_{c0}}{\sqrt{2}}\left( \frac{\xi _{s}}{d_{s}}%
\right) \left( \frac{\xi _{s}}{\xi _{f}}\right) .$ Note that for the
parallel orientation case, the critical temperature must be the same as for
an S/F bilayer with S-layer thickness equals to $d_{s}/2.$ The critical
temperature of S/F bilayers has been recently studied in \cite{Fominov} and
in the limit $d_{s}\ll \xi _{s}$ the expression for $T_{c}$ in \cite{Fominov}
is indeed the same as $\left( \ref{temperature critique}\right) $ with $\tau
=\tau ^{P}$ and $d_{s}$ replaced by $d_{s}/2.$ In the limit of infinite
F-layers $\left( d_{f}\rightarrow \infty \right) $ and infinite transparency
of the interfaces $\left( \gamma _{B}\rightarrow 0\right) ,$ the expression $%
\left( \ref{temperature critique}\right) $ reproduces the results for $T_{c}$
found previously in \cite{Ryzhanova99,baladie2}. If the proximity effect is
weak, the parameter $\tau _{0}^{-1}$ goes to zero. Expanding the Digamma
function about $1/2$ yields the following result in this limit

\begin{equation}
T_{c}^{A}=T_{c}^{P}=T_{c0}-\frac{\pi }{2}\tau ^{A}\left( d_{f}\right) ^{-1}.
\end{equation}
Thus for a weak proximity effect, the shift of the transition temperature is
a linear function of $\tau _{0}^{-1}$ (we find here the same result as in
the study of a superconducting alloy with magnetic impurities) and the
difference Bautzen the critical temperatures of parallel and anti-parallel
orientation appears only at the order $\tau _{0}^{-3}.$

The different kinds of obtained $T_{c}\left( d_{f}\right) $ curves,
depending on parameters of the trilayers are presented in Fig. $2$ for
illustration. We plot several curves for various values of $\gamma _{B}$
assuming that the parameter $\pi T_{c0}\tau _{0}$ is constant and equal to
one. We may notice four characteristic types of $T_{c}\left( d_{f}\right) $
behavior. The first one Fig. $2(a)$, at small interface transparency, $T_{c}$
decays slightly non-monotonously to a finite value and the critical
temperature difference between both orientations is very small. The decay
presents a minimum at a particular value of $d_{f}$ of the order of
magnitude of $\xi _{f}$. The second one Fig. $2(b)$, at moderate interface
transparency, $T_{c}$ exhibits a reentrant behavior, it means that the
superconductivity vanishes in a certain interval of $d_{f}.$ For special
values of the parameter $\gamma _{B}$ the reentrant behavior can be observed
only for the parallel orientation. The reentrance of the superconductivity
has been observed recently in Fe/V/Fe trilayers with parallel orientation of
the ferromagnetic moments by Tagirov et al, see \cite{Tagirov2002}. The
third one Fig. $2(c)$, at moderately high interface transparency, the
critical temperature decays monotonously and vanishes at finite value of $%
d_{f}.$ The last type of $T_{c}(d_{f})$ behavior Fig. $2(d)$ is observed at
really high interface transparency and rather thin F-layers with parallel
orientation. Under these conditions the phase transition between the normal
and the superconducting state presents a triple point at which the
transition switch to the first order one.

In order to observe experimentally a significative spin-valve effect, it is
crucial to choose the right materials and thicknesses of superconductor and
F-layers to maximize $\Delta T_{c}=T_{c}^{P}-T_{c}^{A}.$ Equation $\left( 
\ref{temperature critique}\right) $ shows that the important parameters are $%
\gamma _{B},$ $d_{f}$ and $\tau _{0}^{-1}.$ The value of $\tau _{0}^{-1}$ is
directly related to the choice of the superconductor and of the ferromagnet
since it is proportional to $\gamma $ the ratio of the conductivities. This
parameter does not play the crucial role in the spin-valve effect and a
choice of $\tau _{0}^{-1}$ around one should permit an easy observation of
the effect. The choice of the thickness of the F-layer can be rather
important, as shown by the curves $T_{c}\left( d_{f}\right) .$ Indeed, due
to the additional boundary condition at the interface between the
ferromagnet and the vacuum, there are some interferences between incoming
and reflected Copper pairs in the F-layer. Depending on the value of $d_{f},$
these interferences can be destructive or constructive, leading to a maximum
or a minimum of $\Delta T_{c}.$ Finally, the curves $T_{c}\left(
d_{f}\right) ,$ see Fig. $2$, show that the key factor of the spin-valve
effect is the transparency of the interface. For values of $\gamma _{B}$
around one the effect can be easily observed whereas, if $\gamma _{B}$ is an
order of magnitude stronger the effect almost disappears.

In our case, both F-layers have the same thickness. The generalization to
the case of F-layers of arbitrary thickness $d_{f1\text{ }}$(for $x<0$) and $%
d_{f2}$ (for $x>0$) is straightforward using $\left( \ref{dérivée de f dans
supra}\right) .$ In the parallel case we have to make the substitution $\tau
^{P}\left( d_{f}\right) ^{-1}\rightarrow \tau ^{P}\left( d_{f1}\right)
^{-1}+\tau ^{P}\left( d_{f2}\right) ^{-1}$ and in the anti-parallel case we
have to make the substitution $\tau ^{A}\left( d_{f}\right) ^{-1}\rightarrow
\left( \tau ^{P}\left( d_{f1}\right) ^{-1}\right) ^{\ast }+\tau ^{P}\left(
d_{f2}\right) ^{-1}.$

\section{Thermodynamic properties of the structure}

In this section, we will consider the temperature dependance of the
superconducting order parameter and the superconducting condensation energy
in F/S/F systems. For simplicity, we concentrate on the case of F-layers of
thickness $d_{f}\gg \xi _{f},$ which corresponds in practice to $d_{f}\geq
50 $\AA . Using the classical parametrization of the Usadel equation by $%
F=\sin \theta $ and $G=\cos \theta ,$ we may easily find the complex angle $%
\theta (x)$ in our limit of infinite F-layers for parallel orientation \cite
{baladie} 
\begin{eqnarray}
\theta _{f}^{P}\left( x>d_{s}/2\right) &=&4\arctan \left[ \tan \left( \theta
_{0}^{P}/4\right) \exp \left( -k_{n}\left( x-d_{s}/2\right) \right) \right] ,
\nonumber \\
\theta _{f}^{P}\left( x<-d_{s}/2\right) &=&4\arctan \left[ \tan \left(
\theta _{0}^{P}/4\right) \exp \left( k_{n}\left( x+d_{s}/2\right) \right) %
\right] ,
\end{eqnarray}
and for the anti-parallel one

\begin{eqnarray}
\theta _{f}^{A}\left( x>d_{s}/2\right) &=&4\arctan \left[ \tan \left( \theta
_{0}^{A}/4\right) \exp \left( -k_{n}\left( x-d_{s}/2\right) \right) \right] ,
\nonumber \\
\theta _{f}^{A}\left( x<-d_{s}/2\right) &=&4\arctan \left[ \tan \left(
\theta _{0}^{A}/4\right) \exp \left( k_{n}^{\ast }\left( x+d_{s}/2\right)
\right) \right] ,
\end{eqnarray}
where $\theta _{0}$ is the complex angle describing the superconducting
order parameter in F-layer at S/F boundary. Note that we have assumed in the
previous equations that $\omega $ is positive (the case $\omega <0$ is
obtained by the substitution $k_{n}\rightarrow k_{n}^{\ast }$). These
solutions give us immediately the ratios $\left( F_{f}^{\prime
}/F_{f}\right) _{\pm d_{s}/2}$ and so via the boundary conditions $\left( 
\ref{fprime condition aux limites}\right) ,$ the ratios $\left(
F_{s}^{\prime }/F_{s}\right) _{\pm d_{s}/2}$ which determines the
pair-breaking parameter in the Usadel equations for the S-layer

\begin{eqnarray}
\left( F_{s}^{\prime }/F_{s}\right) _{d_{s}/2} &=&-\frac{\gamma k_{n}\cos
\theta _{0}}{\cos \theta _{0}/2+\widetilde{\gamma _{B}}\cos \theta _{0}}, 
\nonumber \\
\left( F_{s}^{\prime }/F_{s}\right) _{-d_{s}/2}^{P} &=&-\left( F_{s}^{\prime
}/F_{s}\right) _{d_{s}/2}, \\
\left( F_{s}^{\prime }/F_{s}\right) _{-d_{s}/2}^{A} &=&\frac{\gamma
k_{n}^{\ast }\cos \theta _{0}^{A}}{\cos \theta _{0}^{A}/2+\widetilde{\gamma
_{B}}^{\ast }\cos \theta _{0}^{A}}.  \nonumber
\end{eqnarray}
Using $\left( \ref{frpimes sur fs}\right) ,$ we can deduce the coefficient $%
\beta $ for the effective Usadel equations for both orientations $\left(
\omega >0\right) $

\begin{eqnarray}
\beta ^{P} &=&-\frac{2\gamma k_{n}\cos \theta _{0}^{P}}{d_{s}\left( \cos
\theta _{0}^{P}/2+\widetilde{\gamma _{B}}\cos \theta _{0}^{P}\right) },
\label{Usadel Parallèle} \\
\beta ^{A} &=&-\left( \frac{\gamma k_{n}\cos \theta _{0}^{A}}{d_{s}\left(
\cos \theta _{0}^{A}/2+\widetilde{\gamma _{B}}\cos \theta _{0}^{A}\right) }%
+c.c.\right) .  \label{Usadel anti-parallèle}
\end{eqnarray}
The equation $\left( \ref{Usadel et beta}\right) $ with\ $\left( \ref{Usadel
Parallèle},\ref{Usadel anti-parallèle}\right) $ gives, in an implicit form,
the angle $\theta $ (and so the Usadel Green) functions as a function of the
Matsubara frequencies and the superconducting order parameter $\Delta .$
Together with the self-consistency equation $\left( \ref{Auto-cohérence
générale}\right) ,$ this permits in principle to find the dependance of the
superconducting order parameter on temperature and all the thermodynamics of
the F/S/F system. Below, we will discuss two limiting cases which can be
handled analytically: the low temperature limit and temperatures close to $%
T_{c}.$

\subsection{Low temperature behavior}

When the temperature goes to zero, we may substitute the integration by a
summation over Matsubara frequencies $\left( \pi T\sum \rightarrow \int
d\omega \right) $ in the Usadel self-consistency equation for the order
parameter $\left( \ref{Auto-cohérence générale}\right) $

\begin{equation}
\Delta =\lambda N(0)\int_{-\omega _{D}}^{\omega _{D}}F\left( \omega \right)
d\omega =\lambda N(0)\int_{-\omega _{D}}^{\omega _{D}}\sin \theta d\omega ,
\label{Auto-cohérence intégrale}
\end{equation}
where $\omega _{D}$ of the order of magnitude of the Debye frequency is the
usual cut-off in the BCS model (it will not enter in the final expressions)
and $\lambda $ is the BCS coupling constant. The integration over the
Matsubara frequencies can be performed analytically when the transparency of
the S/F interface is small and when it goes to infinity.

\subsubsection{The high transparency limit}

In the high transparency limit $\left( \gamma _{B}\rightarrow 0\right) ,$
the angle $\theta ,$ characterizing superconductivity in the S-layer, is the
same as at the S/F interface, i.e. $\theta _{0}$, see $\left( \ref
{Conditions aux limites}\right) $. Thus the Usadel equations, for parallel
and anti-parallel cases become $\left( \omega >0\right) $

\begin{eqnarray}
\left( \omega +\frac{2\left( 1+i\right) \tau _{0}^{-1}}{\cos \theta
_{0}^{P}/2}\right) \sin \theta _{0}^{P} &=&\Delta \cos \theta _{0}^{P},
\label{usadel trasparence granpe parallele} \\
\left( \omega +\frac{\left( 1+i\right) \tau _{0}^{-1}}{\cos \theta _{0}^{A}/2%
}+c.c.\right) \sin \theta _{0}^{A} &=&\Delta \cos \theta _{0}^{A}.
\label{usadel trasparence granpe anti-parallele}
\end{eqnarray}
Note that these equations are quite different from the corresponding
expressions found in the case of a superconductor with magnetic impurities 
\cite{Abrikosov-Gorkov} and the analogy which worked for $T_{c}$ is no
longer applicable. Let us first consider the parallel case. The integral $%
\left( \ref{Auto-cohérence intégrale}\right) $ can be performed analytically
by changing the integration over $\omega $ by integration over $\theta $

\begin{equation}
\Delta ^{P}=\lambda N(0)\int_{\Delta ^{P}/\omega _{D}}^{\widetilde{\theta }%
^{P}}\left[ \frac{\Delta ^{P}}{\sin ^{2}\theta }+2\tau _{0}^{-1}\left(
1+i\right) \frac{\sin \theta /2}{\cos ^{2}\theta /2}\right] \sin \theta
d\theta +c.c.,  \label{auto-cohérence theta}
\end{equation}
where $\tau _{0}^{-1}$ is given by $\frac{\gamma T_{c0}}{\sqrt{2}}\left( 
\frac{\xi _{s}}{d_{s}}\right) \left( \frac{\xi _{s}}{\xi _{f}}\right) ,$ and 
$\widetilde{\theta }^{P}$ is the solution of equation

\begin{equation}
\Delta ^{P}\cos \widetilde{\theta }^{P}=4\tau _{0}^{-1}\left( 1+i\right)
\sin \left( \widetilde{\theta }^{P}/2\right) .  \label{thetap définition}
\end{equation}
In the absence of the F-layers and at zero temperature, the order parameter $%
\Delta _{0}$ verifies $\left( \ref{auto-cohérence theta}\right) $ with $\tau
_{0}^{-1}=0,$ i.e.

\begin{equation}
\Delta _{0}=2\lambda N(0)\int_{\Delta _{0}/\omega _{D}}^{\pi /2}\frac{\Delta
_{0}}{\sin \theta }d\theta =2\lambda N(0)\Delta _{0}\left( -\ln \left( \tan 
\frac{\Delta _{0}}{2\omega _{D}}\right) \right) .
\label{constante de couplage2}
\end{equation}

Combining $\left( \ref{auto-cohérence theta},\ref{constante de couplage2}%
\right) ,$ we may eliminate the diverging terms when $\theta $ goes to zero
and finally, performing the remaining integration, we obtain the following
explicit relation for the ratio$\ \Delta ^{P}/\Delta _{0}$

\begin{equation}
\ln \left( \frac{\Delta ^{P}}{\Delta _{0}}\right) =%
\mathop{\rm Re}%
\left\{ \ln \tan \left( \widetilde{\theta }^{P}/2\right) +4\tau _{0}^{-1}%
\frac{(1+i)}{\Delta ^{P}}\left[ \ln \tan \left( \frac{\widetilde{\theta }%
^{P}+\pi }{4}\right) -\sin \left( \widetilde{\theta }^{P}/2\right) \right]
\right\} .  \label{Delta parallèle}
\end{equation}
Performing the same kind of calculation in the anti-parallel case we have
for the ratio $\Delta ^{A}/\Delta _{0}$

\begin{equation}
\ln \left( \frac{\Delta ^{A}}{\Delta _{0}}\right) =\ln \tan \left( 
\widetilde{\theta }^{A}/2\right) +\frac{4\tau _{0}^{-1}}{\Delta ^{A}}\left[
\ln \tan \left( \frac{\widetilde{\theta }^{A}+\pi }{4}\right) -\sin \left( 
\widetilde{\theta }^{A}/2\right) \right] ,  \label{Delta anti-parallèle}
\end{equation}
where $\widetilde{\theta }^{A}$ is the solution of

\begin{equation}
\Delta ^{A}\cos \widetilde{\theta }^{A}=4\tau _{0}^{-1}\sin \left( 
\widetilde{\theta }^{A}/2\right) .  \label{thetaa définition}
\end{equation}

The density of states for one direction of spin is given by $N_{\uparrow
}\left( \omega \right) =\frac{1}{2}N(0)%
\mathop{\rm Re}%
\left( G\left( \omega \rightarrow i\omega \right) \right) ,$ where $N(0)$ is
the total density of state in the normal state$.$ Considering the limit $%
\omega =0,$ this relation becomes $N_{\uparrow }\left( \omega \right) =\frac{%
1}{2}N(0)%
\mathop{\rm Re}%
\left( \cos \widetilde{\theta }^{A,P}\right) ,$ where $\widetilde{\theta }%
^{P}$ is given by $\left( \ref{thetap définition}\right) $ and $\widetilde{%
\theta }^{A}$ by $\left( \ref{thetaa définition}\right) .$ An analytical
study of $\left( \ref{thetap définition},\ref{thetaa définition}\right) $
shows that the real part of the solutions $\widetilde{\theta }^{A,P}$ always
exists, thus $N_{\uparrow }\left( \omega \right) $ is finite at $\omega =0.$
As a result, at low temperatures and in both configurations, the
superconductivity in F/S/F systems should be a gapless one.

In Fig. $3$, we have plotted the order parameter in both parallel and
anti-parallel case as a function of the pair-breaking parameter $\left(
\Delta _{0}\tau _{0}\right) ^{-1}.$ At small exchange field or at small
conductivities ratio, there is almost no difference between $\Delta ^{P}$
and $\Delta ^{A}$ and their evolution with $\left( \Delta _{0}\tau
_{0}\right) ^{-1}$ is linear as in the case of superconducting alloys
containing magnetic impurities, however the overall behavior in the whole
temperature region is different. Naturally, the superconducting order
parameter is always larger in the anti-parallel case due to the partial
compensation of the exchange field effect.

The thermodynamic potential (per unit area) for both orientations can be
found by integrating $\left( \ref{Delta parallèle},\ref{Delta anti-parallèle}%
\right) $

\begin{eqnarray}
\Omega ^{P}\left( h,\Delta \right) &=&d_{s}N(0)\left( \frac{\Delta ^{2}}{2}%
\ln \frac{\Delta ^{2}}{e\Delta _{0}^{2}}-\tau _{0}^{-2}%
\mathop{\rm Re}%
\left[ if(X^{P})\right] \right) ,  \label{Potentiel thermo parallèle} \\
\Omega ^{A}\left( h,\Delta \right) &=&d_{s}N(0)\left( \frac{\Delta ^{2}}{2}%
\ln \frac{\Delta ^{2}}{e\Delta _{0}^{2}}-\tau _{0}^{-2}f(X^{A})\right) ,
\label{Potentiel thermo anti-parallèle}
\end{eqnarray}
where the function $f(X)$ is defined by

\begin{equation}
\frac{1}{2\left( 2\left( X\right) ^{2}-1\right) ^{2}}\left[ 
\begin{array}{c}
-\left( X+1\right) ^{2}\left( 2X-1\right) ^{2}\ln \left( 1+X\right) -\left(
X-1\right) ^{2}\left( 2X+1\right) ^{2}\ln \left( 1-X\right)  \\ 
+2X^{2}\ln X+6X^{2}\left( 2X^{2}-1\right) 
\end{array}
\right] ,
\end{equation}
and while $X^{A,P}=\sin \left( \widetilde{\theta }^{A,P}/2\right) $.
Minimizing $\left( \ref{Potentiel thermo parallèle},\ref{Potentiel thermo
anti-parallèle}\right) $ in respect to the order parameter at fixed exchange
field gives back the self-consistency equations $\left( \ref{Delta parallèle}%
,\ref{Delta anti-parallèle}\right) $ determining $\Delta \left( h\right) .$
Keeping in mind the fact that the free energy $F$ of the system is equal to
the thermodynamic potential when the order parameter is minimized, we have
determined the difference of free energy between the parallel and the
anti-parallel configuration $F^{P}-F^{A}=\Omega ^{P}\left( h,\Delta
^{P}\right) -\Omega ^{A}\left( h,\Delta ^{A}\right) .$ The analysis of $%
\left( \ref{temperature critique}\right) $ in the case of infinite F-layers
and high transparency of the interfaces immediately shows that the
superconducting transition temperature is going to zero for 
\begin{eqnarray}
\left( \Delta _{0}\tau _{0}\right) ^{-1} &=&0.25\text{ in the parallel case,}
\\
\left( \Delta _{0}\tau _{0}\right) ^{-1} &=&0.175\text{ in the anti-parallel
case,}
\end{eqnarray}
These values naturally correspond to gaps vanishing in Fig. $3.$ As a
result, $F^{P}$ is equal to zero for $\left( \Delta _{0}\tau _{0}\right)
^{-1}>0.25$ and $F^{A}$ for $\left( \Delta _{0}\tau _{0}\right) ^{-1}>0.175.$
In Fig. $4$\ , we have plotted the normalized expression of $\left(
F^{P}-F^{A}\right) $, by the free energy in the anti-parallel configuration,
as a function of the parameter $\left( \Delta _{0}\tau _{0}\right) ^{-1}.$
The expression $\left( F^{P}-F^{A}\right) $ is always positive, in
conclusion, the anti-parallel configuration is always more stable than the
parallel configuration.

\subsubsection{The low transparency limit}

In this limit, an expansion of equations $\left( \ref{Usadel Parallèle},\ref
{Usadel anti-parallèle}\right) $ with respect to $1/\gamma _{B}$ can be
made. In the limit of low transparency $\left( \gamma _{B}\rightarrow \infty
\right) $, the order parameter in the F-layer almost completely disappears,
thus the angle $\theta _{0}$ describing the superconducting order parameter
in F-layer at S/F boundary is\ small $\left( \theta _{0}\ll 1\right) $. So,
with the help of the boundary conditions $\left( \ref{Conditions aux limites}%
\right) ,$ we find that the angle $\theta ,$ characterizing
superconductivity in the S-layer, is given in this limit by $\sin \theta
=\theta _{0}\widetilde{\gamma _{B}}/2.$ With this relation and $\left( \ref
{frpimes sur fs}\right) ,$ we can easily find the expression of the
coefficient $\beta $ in both configurations

\begin{eqnarray}
\beta ^{P} &=&-\frac{\gamma }{\xi _{f}d_{s}\gamma _{B}}\left( 1-\widetilde{%
\gamma _{B}}^{-1}+2\widetilde{\gamma _{B}}^{-2}\right) , \\
\beta ^{AP} &=&\left( \beta ^{P}+\beta ^{P\ast }\right) /2.
\end{eqnarray}

The stability of both parallel and anti-parallel configurations of the FSF
trilayer in the low transparency limit can also be studied by performing the
integration over $\omega $ in the Usadel self-consistency equation $\left( 
\ref{Auto-cohérence intégrale}\right) $. We only gives here the results of
the corresponding calculations

\begin{eqnarray}
\ln \left( \frac{\Delta ^{P}}{\Delta _{0}}\right) &=&-\sqrt{\frac{2}{\Delta
^{P}\tau _{0}\gamma _{B}}}\left( 1-\frac{1}{2\sqrt{2}\gamma _{B}}\right) +%
\frac{\pi }{4\Delta ^{P}\tau _{0}\gamma _{B}}\left( 1-\frac{1}{\sqrt{2}%
\gamma _{B}}\right) ,  \label{deltaP fct gamma} \\
\ln \left( \frac{\Delta ^{A}}{\Delta _{0}}\right) &=&-\sqrt{\frac{2}{\Delta
^{A}\tau _{0}\gamma _{B}}}\left( 1-\frac{1}{2\sqrt{2}\gamma _{B}}-\frac{1}{%
16\gamma _{B}^{2}}\right) +\frac{\pi }{4\Delta ^{A}\tau _{0}\gamma _{B}}%
\left( 1-\frac{1}{\sqrt{2}\gamma _{B}}\right) .  \label{deltaAP fct de gamma}
\end{eqnarray}
Following the method presented in the previous paragraph, we obtain the
expression for the thermodynamic potential

\begin{equation}
\Omega ^{A,P}\left( h,\Delta \right) =d_{s}N(0)\left( \frac{\Delta ^{2}}{2}%
\ln \frac{\Delta ^{2}}{e\Delta _{0}^{2}}+a^{A,P}\Delta ^{3/2}-b^{A,P}\Delta
\right) ,  \label{Potentiel thermo tranparence faible}
\end{equation}
where the coefficients $a^{P}=\frac{4}{3}\sqrt{\frac{2}{\tau _{0}\gamma _{B}}%
}\left( 1-\frac{1}{2\sqrt{2}\gamma _{B}}\right) ,$ $a^{A}=\frac{4}{3}\sqrt{%
\frac{2}{\tau _{0}\gamma _{B}}}\left( 1-\frac{1}{2\sqrt{2}\gamma _{B}}-\frac{%
1}{16\gamma _{B}^{2}}\right) $ and $b^{A,P}=\frac{\pi }{2\tau _{0}\gamma _{B}%
}\left( 1-\frac{1}{\sqrt{2}\gamma _{B}}\right) .$ The term containing $%
\gamma _{B}^{-2}$ in $a^{A}$ contributes to the stabilization of the
anti-parallel configuration compared to the parallel configuration.
Although, as it follows from $\left( \ref{deltaP fct gamma},\ref{deltaAP fct
de gamma},\ref{Potentiel thermo tranparence faible}\right) $, the
orientation dependent relative variation of the order parameter and
condensation energy is very small 
\begin{equation}
\frac{\Delta ^{P}-\Delta ^{A}}{\Delta _{0}}\sim \frac{F^{P}-F^{A}}{F_{0}}%
\sim \frac{\gamma _{B}^{-5/2}}{\sqrt{\Delta _{0}\tau _{0}}}.
\end{equation}

\subsection{Free energy, entropy and specific heat of the trilayer close to $%
T_{c}$}

At the transition temperature, the order parameter $\Delta $ goes to zero
and the Green functions $F$ and $G$ go respectively to $0$ and $sgn\left(
\omega \right) .$ In the limit of high S/F interfaces transparency, we may
use $\left( \ref{usadel trasparence granpe parallele},\ref{usadel
trasparence granpe anti-parallele}\right) $ and develop all the quantities
around $T_{c}$ to obtain an expansion of $F$ in powers of $\Delta $

\begin{equation}
F=\frac{\Delta }{\left| \omega \right| +\epsilon \left( \omega \right) }-%
\frac{\Delta ^{3}}{2\left( \left| \omega \right| +\epsilon \left( \omega
\right) \right) ^{3}}-\frac{\Delta ^{3}\epsilon \left( \omega \right) }{%
8\left( \left| \omega \right| +\epsilon \left( \omega \right) \right) ^{4}}%
+o\left( \Delta ^{5}\right) ,  \label{auto-cohérence pour Tc}
\end{equation}
where for the parallel case $\epsilon =\epsilon ^{P}\left( \omega \right) =%
\frac{2\left( 1+i\right) \tau _{0}^{-1}}{\cos \theta _{0}^{P}/2}$ while in
the parallel case $\epsilon =\epsilon ^{A}\left( \omega \right) =\frac{%
\epsilon ^{P}\left( \omega \right) +\epsilon ^{P}\left( \omega \right)
^{\ast }}{2}.$ Thus, using expression $\left( \ref{auto-cohérence pour Tc}%
\right) $ and the self-consistency equation, we may directly obtain the
dependance of the order parameter with the temperature. In the anti-parallel
case we have

\begin{equation}
-\ln \left( \frac{T}{T_{c0}}\right) =\Psi \left( \frac{1}{2}+\frac{1}{\pi
\tau _{0}T}\right) -\Psi \left( \frac{1}{2}\right) +\left( \frac{\Delta }{%
2\pi T}\right) ^{2}g_{1}\left( \frac{1}{\pi \tau _{0}T}\right) ,
\label{Paramètre d'ordre AP}
\end{equation}
and in the parallel case

\begin{equation}
-\ln \left( \frac{T}{T_{c0}}\right) =%
\mathop{\rm Re}%
\Psi \left( \frac{1}{2}+\frac{\left( 1+i\right) }{\pi \tau _{0}T}\right)
-\Psi \left( \frac{1}{2}\right) +\left( \frac{\Delta }{2\pi T}\right) ^{2}%
\mathop{\rm Re}%
g_{1}\left[ \frac{\left( 1+i\right) }{\pi \tau _{0}T}\right] ,
\label{Paramètre d'ordre P}
\end{equation}
where the function $g_{1}(x)=-\frac{1}{4}\Psi ^{(2)}\left( \frac{1}{2}%
+x\right) +\frac{x}{48}\Psi ^{(3)}\left( \frac{1}{2}+x\right) .$ It's
important to note that the function $g_{1}(x)$ is positive for all values of 
$\left( \pi \tau _{0}T\right) ^{-1},$ so the superconducting phase
transition is always a second order one for $d_{f}\gg \xi _{f}$. The
transition temperature of the superconductor in contact with the F-layers is
determined by putting $\Delta =0$ in the previous equations and gives back
the results of the previous section and of \cite{Ryzhanova99,baladie2}, in
the limit of large F-layer and large transparency of the interfaces.
Simplifying $\left( \ref{Paramètre d'ordre AP},\ref{Paramètre d'ordre P}%
\right) $ using $\left( \ref{temperature critique}\right) ,$ and defining
the function $g_{2}\left( x\right) =1-x\Psi ^{(1)}\left( \frac{1}{2}%
+x\right) ,$ we get

\begin{eqnarray}
\Delta _{A}^{2} &=&\left( 2\pi T_{c}^{A}\right) ^{2}\frac{g_{2}\left[ \left(
\pi \tau _{0}T_{c}^{A}\right) ^{-1}\right] }{g_{1}\left[ \left( \pi \tau
_{0}T_{c}^{A}\right) ^{-1}\right] }\left( 1-\frac{T}{T_{c}^{A}}\right) ,
\label{delta AP} \\
\Delta _{P}^{2} &=&\left( 2\pi T_{c}^{P}\right) ^{2}\frac{%
\mathop{\rm Re}%
g_{2}\left[ \left( \pi \tau _{0}T_{c}^{P}\right) ^{-1}\right] }{%
\mathop{\rm Re}%
g_{1}\left[ \left( \pi \tau _{0}T_{c}^{P}\right) ^{-1}\right] }\left( 1-%
\frac{T}{T_{c}^{P}}\right) .  \label{delta P}
\end{eqnarray}
This shows that the order parameters increases as $\left( 1-T/T_{c}\right)
^{1/2}$ when the temperature is sufficiently low. The free energy of the
system is simply given by, see \cite{Abrikosov},

\begin{equation}
F_{s}-F_{n}=\Delta F=-\int_{0}^{\lambda }\frac{\Delta ^{2}}{\lambda _{1}^{2}}%
d\lambda _{1}=\int_{0}^{\Delta }\Delta _{1}^{2}\frac{d\left( \frac{1}{%
\lambda }\right) }{d\Delta _{1}}d\Delta _{1}.  \label{energie de Gibbs}
\end{equation}
Using the relation $\delta \left( 1/\lambda \right) =-N(0)\delta
T_{c0}/T_{c0}$, see \cite{Abrikosov}, and equations $\left( \ref{delta AP},%
\ref{delta P}\right) ,$ we obtain $\frac{d\left( \lambda ^{-1}\right) }{%
d\Delta }=-\frac{\Delta N(0)}{2\pi ^{2}T_{c}^{2}}g_{1}\left[ \left( \pi \tau
_{0}T_{c}\right) ^{-1}\right] .$ Thus the calculation of the free energy is
straightforward $\left( \ref{energie de Gibbs}\right) $

\begin{eqnarray}
\Delta F^{A} &=&-2N(0)\pi ^{2}\left( T_{c}^{A}\right) ^{2}\frac{g_{2}^{2}%
\left[ \left( \pi \tau _{0}T_{c}^{A}\right) ^{-1}\right] }{g_{1}\left[
\left( \pi \tau _{0}T_{c}^{A}\right) ^{-1}\right] }\left( 1-\frac{T}{%
T_{c}^{A}}\right) ^{2}, \\
\Delta F^{P} &=&-2N(0)\pi ^{2}\left( T_{c}^{P}\right) ^{2}\frac{\left\{ 
\mathop{\rm Re}%
g_{2}\left[ \left( \pi \tau _{0}T_{c}^{P}\right) ^{-1}\right] \right\} ^{2}}{%
\mathop{\rm Re}%
g_{1}\left[ \left( \pi \tau _{0}T_{c}^{P}\right) ^{-1}\right] }\left( 1-%
\frac{T}{T_{c}^{P}}\right) ^{2}.
\end{eqnarray}
From the above equations, the entropy and the heat capacity are obtained
using $S=-\left( \frac{\partial F}{\partial T}\right) _{h}$ and $C=-T\left( 
\frac{\partial ^{2}F}{\partial T^{2}}\right) _{h}.$ We present the results
obtained for the heat capacity only 
\begin{eqnarray}
\Delta C^{P}\left( T_{c}^{P}\right) &=&4\pi ^{2}N(0)T_{c}^{P}\frac{\left\{ 
\mathop{\rm Re}%
g_{2}\left[ \left( \pi \tau _{0}T_{c}^{P}\right) ^{-1}\right] \right\} ^{2}}{%
\mathop{\rm Re}%
g_{1}\left[ \left( \pi \tau _{0}T_{c}^{P}\right) ^{-1}\right] }, \\
\Delta C^{A}\left( T_{c}^{A}\right) &=&4\pi ^{2}N(0)T_{c}^{A}\frac{g_{2}^{2}%
\left[ \left( \pi \tau _{0}T_{c}^{A}\right) ^{-1}\right] }{g_{1}\left[
\left( \pi \tau _{0}T_{c}^{A}\right) ^{-1}\right] }.
\end{eqnarray}
The jump of the specific heat at the transition decreases monotonically as $%
T_{c}$ decreases (i.e. as the pair-breaking effect of the F-layers
increases). The corresponding results are plotted in Fig. $5$, where the
jump of the specific heat at $T_{c}$ is normalized by the jump of the
specific heat at $T_{c0}$ the critical temperature without any proximity
effect.

\section{Conclusion}

We have considered the properties of F/S/F spin-valve systems and presented
their general theoretical description for the most interesting case of thin
superconducting layer. The spin-valve effect occurs to be very strongly
dependent on the S/F interface transparency. So, to observe it on experiment
it is necessary to chose superconductor-ferromagnet systems with a low
barrier at the interface. The oscillatory-like $T_{c}$ dependance on the
F-layer thickness $d_{f}$ gives the optimum condition of spin-valve
observation for $d_{f}\sim \xi _{f}$ i.e. $10-50$\AA , but the situation
remains qualitatively the same for higher thickness too. The maximum gain in
the superconducting energy corresponds to the anti-parallel configuration,
and this gain may be of the same order of magnitude as the superconducting
condensation energy itself. So we may except that without external applied
field, parallel configuration will be unstable. Therefore, with the decrease
of the temperature below $T_{c},$ the transition from parallel to
anti-parallel configuration may be observed. Since it would depend on the
magnetic cohercitivity force, thin F-layers would be ''a priori'' more
suitable to observe such effect. A very interesting situation can be also
observed when the Curie temperature is lower than the superconducting
critical temperature. In such a case we may except the spontaneous
appearance of the anti-parallel configuration by decreasing the temperature.
It is worth to note that in the case when the domain wall energy is small,
the formation of short length-scale magnetic domains could occur at the
contact of the ferromagnet and the superconductor \cite{Buzdin88,Bergeret}.

In conclusion, the F/S/F trilayer systems reveal strong interferences
between superconducting and magnetic effects. They could be quite
interesting for application as a very small magnetic field may strongly
influence the superconducting characteristics via the spin-valve effect.

\bigskip

We thank C. Baraduc, V. V. Ryazanov, H. Sellier and A. V. Vedyayev for
stimulating discussions. This work was supported by the NATO Collaborative
Linkage Grant No. CLG 978153, the ESF ''vortex'' Programme and the ACI
''supra-nanom\'{e}trique''.

FIG. $1$. Geometry of the F/S/F sandwich. The thickness of the S-layer is $%
d_{s}$ and $d_{f}$ is the thickness of the F-layers.

FIG. $2.$ Characteristic types of $T_{c}\left( d_{f}\right) $ behavior. The
thickness of the F-layer is normalized to the F-layer charactersitic length $%
\xi _{f}.$ The parameter $\pi \tau _{0}T_{c0}$ is choosen constant and equal
to one$.$ The full line corresponds to the anti-parallel case, the dotted
line to the parallel case. One can distinguish four characteristic types of $%
T_{c}\left( d_{f}\right) $ behavior: (a) and (b) monotonic decay to \ $%
T_{c}=0$ with (a) or without (b) switching to a first order transition in
the parallel case, (c) reentrant behavior for the parallel orientation, (d)
nonmonotonic decay to a finite value of $T_{c}.$

FIG. $3.$ The order parameter $\Delta $ normalized by its value in absence
of proximity effect $\Delta _{0}$ in both parallel and anti-parallel case as
a function of the pair breaking parameter $\left( \tau _{0}\Delta
_{0}\right) ^{-1}.$

FIG. $4.$ Normalized value of the difference of free energy between the
parallel and the anti-parallel configurations plotted as a function of the
parameter $\left( \tau _{0}\Delta _{0}\right) ^{-1}.$ For $\left( \tau
_{0}\Delta _{0}\right) ^{-1}\geq 0.175,$ the superconducting transition
temperature is equal to zero in the parallel configuration.

FIG. $5.$ Discontinuity of the specific heat at the critical temperature
versus $T_{c}/T_{c0}.$ The full line corresponds to the anti-parallel case,
the dotted line to the parallel case.


\begin{references}
\bibitem{kuprianov90}  A. I. Buzdin and M. Y. Kuprianov, JETP Lett. {\bf 53, 
}321 (1991).

\bibitem{Radovic91}  Z. Radovic, M. Ledvij, L. Dobrosavljevic-Grujic, A. I.
Buzdin, and J. R. Clem, Phys. Rev. B {\bf 44}, 759 (1991).

\bibitem{Tagirov Physica C}  L. R. Tagirov, Physica C {\bf 307,} 145 (1998).

\bibitem{Proshin}  Y. N. Proshin and M. G. Khusainov, Zh. Eksp. Theor. Fiz. 
{\bf 113}, 1708 (1998) [JETP {\bf 86}, 930 (1998)]; {\bf 116}, 1887 (1999) [%
{\bf 89}, 1021 (1999)].

\bibitem{Jiang85}  J. S. Jiang, D. Davidovic, D. H. Reich, and C. L. Chien,
Phys. Rev. Lett. {\bf 74}, 314 (1985).

\bibitem{Muihge}  T. M\"{u}hge, N. N. Garif'yanov, Yu. V. Goryunov, G. G.
Khaliullin, L. R. Tagirov, K. Westerholt, I. A. Garifullin, and H. Zabel,
Phys. Rev. Lett. {\bf 77,} 1857 (1996).

\bibitem{Lazar}  L. Lazar, K. Westerholt, H. Zabel, L. R. Tagirov, Yu. V.
Goryunov, N. N. Garif'yanov, and I. A. Garifullin, Phys. Rev. B {\bf 61},
3711 (2000).

\bibitem{Panyukov82}  A. I. Buzdin, L. N. Bulaevskii, and S. V. Panyukov,
JETP Lett. {\bf 35}, 178 (1982).

\bibitem{Kuprianov91}  A. I. Buzdin and M. Y. Kuprianov, JETP Lett. {\bf 52, 
}487 (1990).

\bibitem{ryazanov}  V. V. Ryazanov, V. A. Oboznov, A. Yu. Rusanov, A. V.
Veretennikov, A. A. Golubov, and J. Aarts, Phys. Rev. Lett. {\bf 86}, 2427
(2001).

\bibitem{Kontos2}  T. Kontos, M. Aprili, J. Lesueur, F. Genet, B.
Stephanidis, and R. Boursier, cond-mat/0201104 (unpublished).

\bibitem{buzdin00}  A. Buzdin, Phys. Rev. B {\bf 62}, 11377 (2000).

\bibitem{baladie}  I. Baladi\'{e} and A. Buzdin, Phys. Rev. B {\bf 64},
224514 (2001).

\bibitem{kontos}  T. Kontos, M. Aprili, J. Lesueur and X. Grison, Phys. Rev.
Lett. {\bf 86}, 304 (2001).

\bibitem{Ryzhanova99}  A. I. Buzdin, A. V. Vedyayev, and N. V. Ryzhanova,
Europhys. Lett. {\bf 48}, 686 (1999).

\bibitem{baladie2}  I. Baladi\'{e}, A. Buzdin, N. Ryzhanova, and A.
Vedyayev, Phys. Rev. B {\bf 63}, 54518 (2001).

\bibitem{Tagirov}  L. R. Tagirov, Phys. Rev. Lett. {\bf 83}, 2058 (1999).

\bibitem{Usadel70}  L. Usadel, Phys. Rev. Lett. {\bf 95}, 507 (1970).

\bibitem{Kuprianov88}  M. Y. Kuprianov and V. F. Lukichev, Zh. Eksp. Theor.
Fiz. {\bf 94}, 139 (1988) [Sov. Phys. JETP {\bf 67}, 1163 (1988)].

\bibitem{AArts}  J. Aarts, J. M. E. Geers, E. Br\"{u}ck, A. A. Golubov, and
R. Coehoorn, Phys. Rev. B {\bf 56}, 2779 (1997).

\bibitem{Abrikosov}  A. A. Abrikosov, L. P. Gor'kov, and I. E.
Dzyaloshinskii, {\it Methods of Quantum Field Theory in Statistical Physics,}
Dover Publications, INC. (1963).

\bibitem{Fominov}  Ya. V. Fominov, N. M. Chtchelkatchev, and A. A. Golubov,
Phys. Rev. B {\bf 66,} 14507 (2002).

\bibitem{Tagirov2002}  L. R. Tagirov, I. A. Garifullin, N. N. Garif'yanov,
S. Ya. Khlebnikov, D. A. Tikhonov, K. Westerholt and H. Zabel, J. Magn.
Magn. Mater. {\bf 240}, 577 (2002).

\bibitem{Abrikosov-Gorkov}  A. A. Abrikosov and L. P. Gor'kov, Zh. Eksp.
Theor. Fiz. {\bf 39}, 1781 (1960) [Sov. Phys. JETP {\bf 12}, 1243 (1961)].

\bibitem{Buzdin88}  A. I. Buzdin and L. N. Bulaevskii, Zh. Eksp. Teor. Fiz. 
{\bf 94}, 256 (1988) [Sov. Phys. JETP {\bf 67}, 576 (1988)].

\bibitem{Bergeret}  F. S. Bergeret, K. B. Efetov, and A. I. Larkin, Phys.
Rev. B {\bf 62}, 11872 (2000).
\end{references}
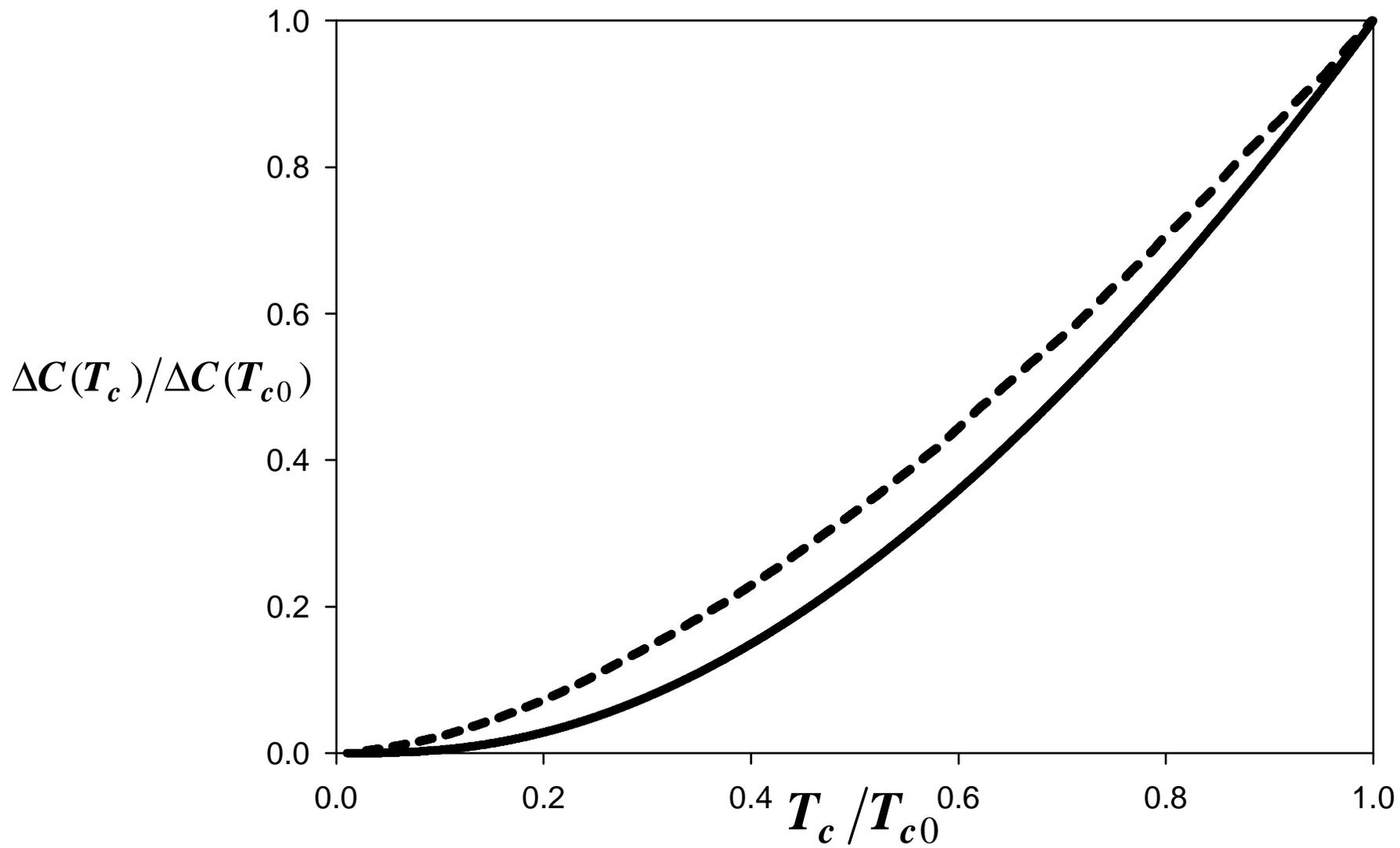
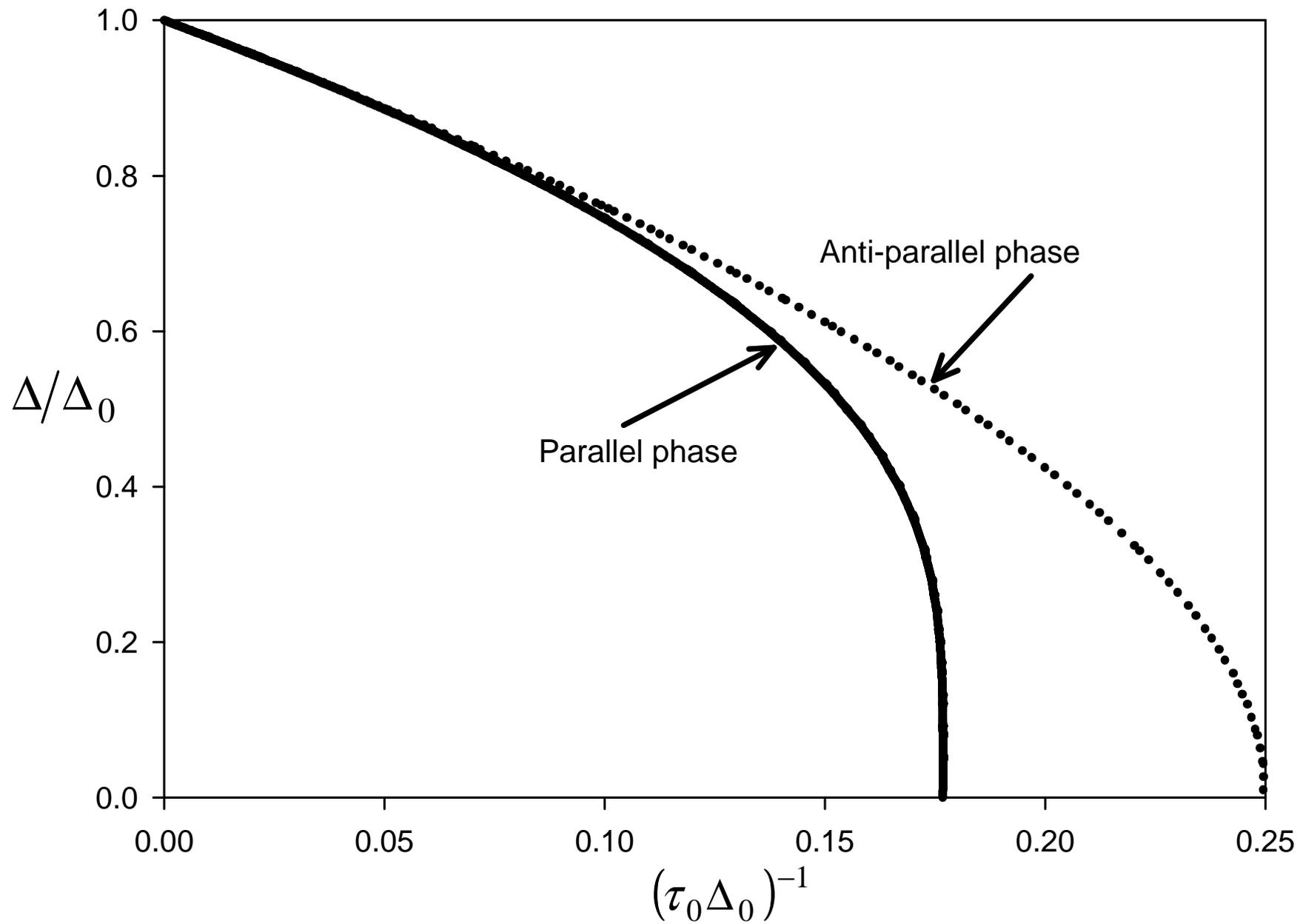
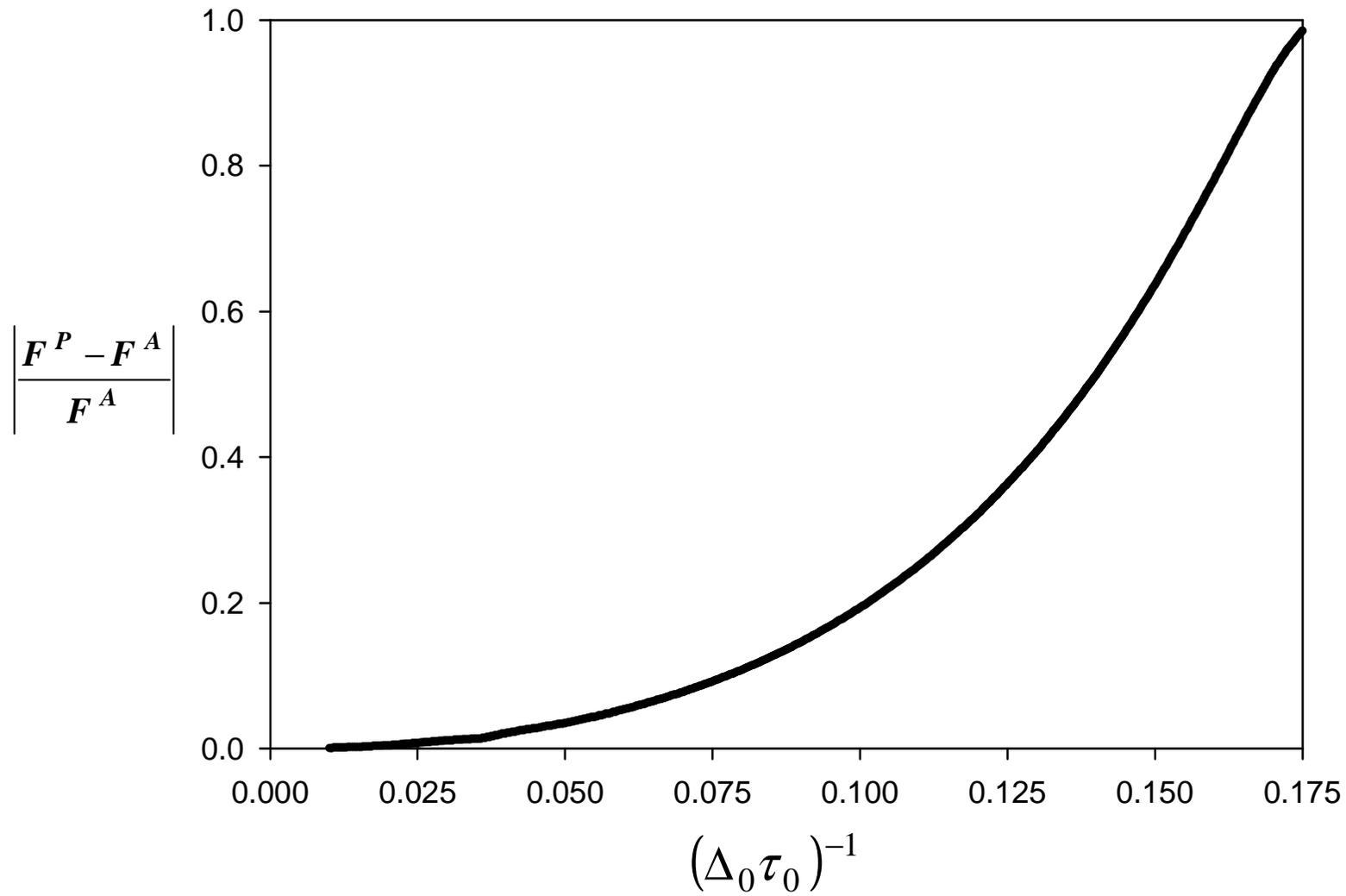
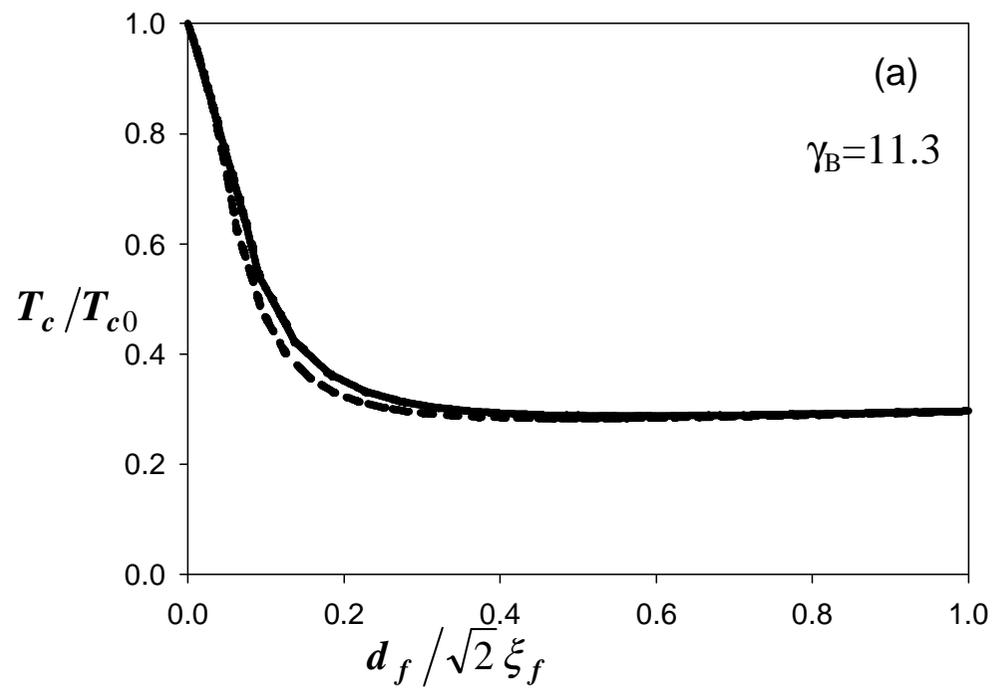
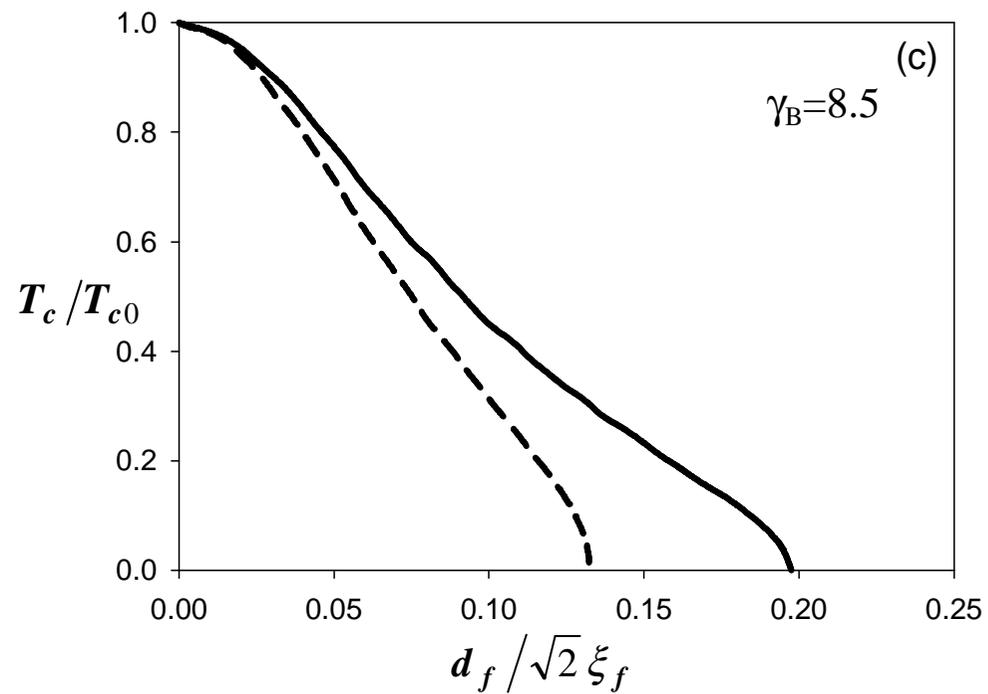
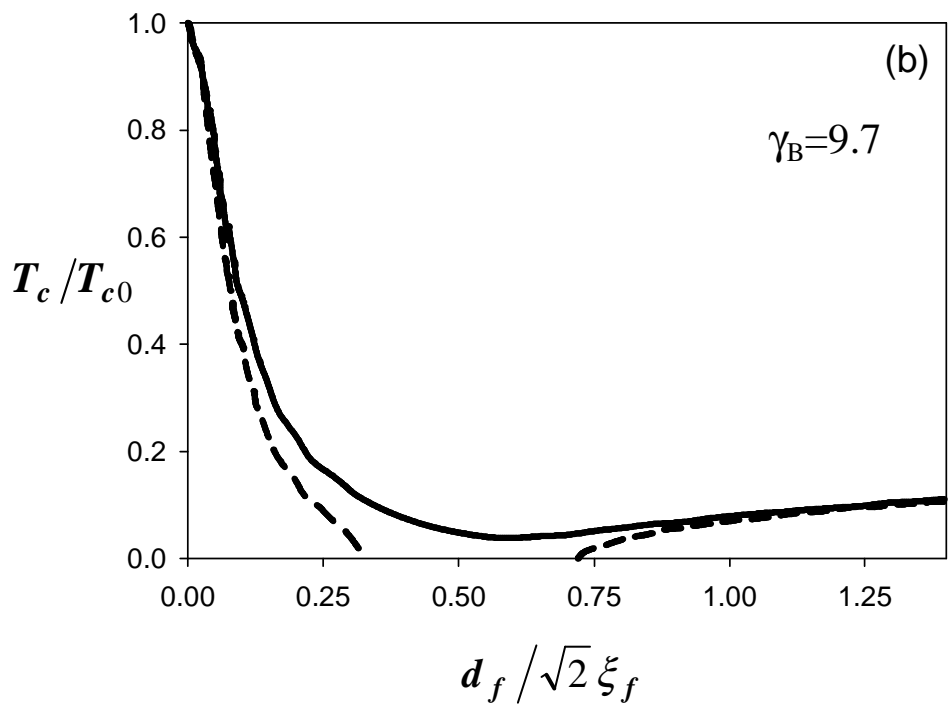
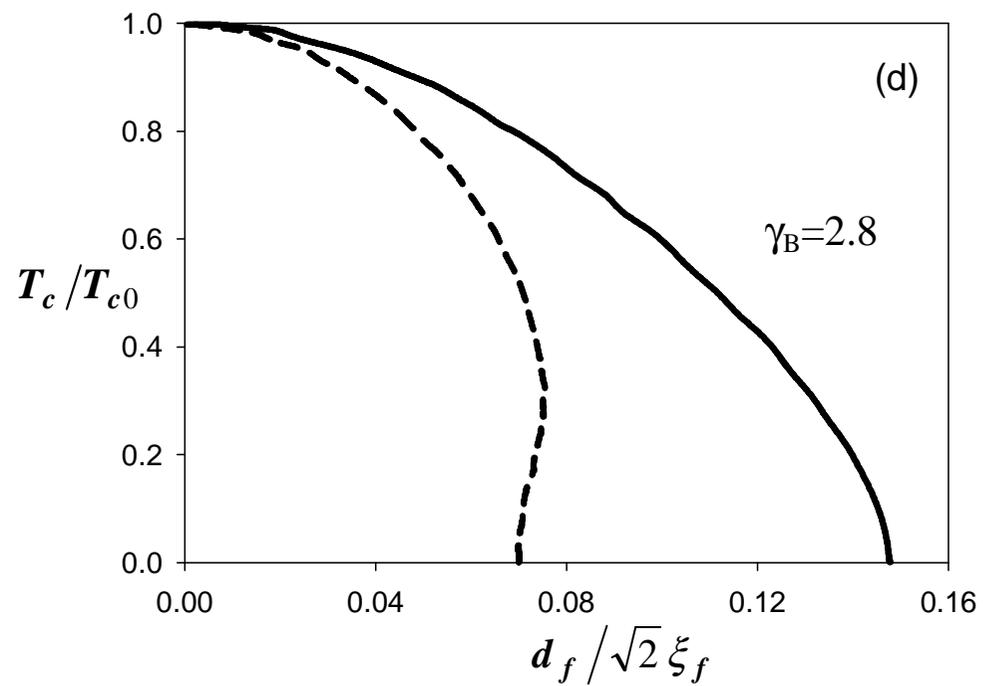
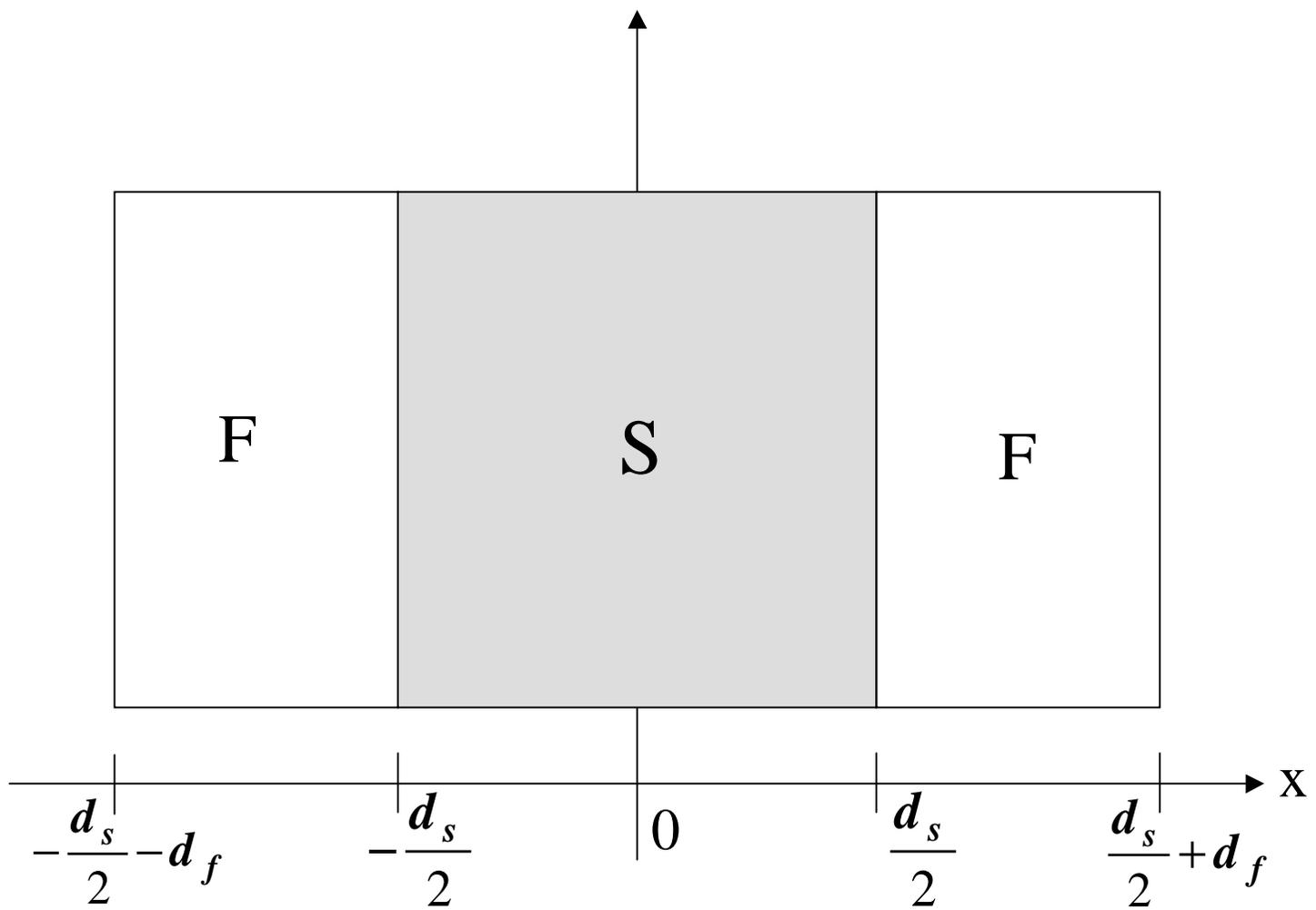
\end{document}